\documentclass[letter,A4paper]{aa} 
\usepackage{natbib}
\usepackage{lscape}
\usepackage{rotating}
\usepackage{graphicx}
\usepackage{amsmath}
\usepackage{enumerate}
\usepackage{txfonts}
\begin{document}
  \title{The timescale of low-mass proto-helium white dwarf evolution 
        }
  
   \author{A. G. Istrate
          \inst{1}\fnmsep\thanks{e-mail: aistrate@astro.uni-bonn.de},
          T. M. Tauris\inst{1,2}
          N. Langer\inst{1}
           \and
          J. Antoniadis\inst{2}
          }

   \authorrunning{Istrate, Tauris, Langer \& Antoniadis}
   \titlerunning{The timescale of low-mass proto-He WD evolution}

   \institute{Argelander-Institut f\"ur Astronomie, Universit\"at Bonn, Auf
              dem H\"ugel 71, 53121 Bonn, Germany
         \and
              Max-Planck-Institut f\"ur Radioastronomie, 
              Auf dem H\"ugel 69, 53121 Bonn, Germany 
              }

   \date{Received July 25, 2014; accepted October 17, 2014}

 
  \abstract
   {A large number of low-mass ($<0.20\;M_{\odot}$) helium white dwarfs (He~WDs) have recently been discovered. The majority of these 
    are orbiting another WD or a millisecond pulsar (MSP) in a close binary system; a few examples are found to show pulsations or to have 
    a main-sequence star companion. There appear to be discrepancies between the current theoretical modelling of such low-mass He~WDs 
    and a number of key observed cases, indicating that their formation scenario  yet remains to be fully understood. }
   {Here we investigate the formation of detached proto-He~WDs in close-orbit low-mass X-ray binaries (LMXBs).
    Our prime focus is to examine the thermal evolution and the contraction phase towards the WD cooling track
    and investigate how this evolution depends on the WD mass.
    Our calculations are then compared to the most recent observational data.} 
   {Numerical calculations with a detailed stellar evolution code were used to trace the mass-transfer phase 
    in a large number of close-orbit LMXBs with different initial values of donor star mass, neutron star mass, orbital period,
    and strength of magnetic braking. 
    Subsequently, we followed the evolution of the detached low-mass proto-He~WDs, including stages with residual shell 
    hydrogen burning and vigorous flashes caused by unstable CNO burning.}
    {We find that the time between Roche-lobe detachment until the low-mass proto-He~WD reaches the WD cooling track is typically 
    $\Delta t_{\rm proto}=0.5-2\;{\rm Gyr}$, depending systematically on the WD mass and therefore on its luminosity. 
    The lowest WD mass for developing shell flashes is $\sim\!0.21\;M_{\odot}$ for progenitor stars of mass $M_2 \le 1.5\;M_{\odot}$
    (and $\sim\!0.18\;M_{\odot}$ for $M_2=1.6\;M_{\odot}$).}
   {The long timescale of low-mass proto-He~WD evolution can explain a number of recent observations,  
    including some MSP systems hosting He~WD companions with very low surface gravities and high effective temperatures.  
    We find no evidence for $\Delta t_{\rm proto}$ to depend on the occurrence of flashes and thus question the suggested dichotomy
    in thermal evolution of proto-WDs.
    }


   \maketitle
%

\section{Introduction}\label{sec:intro}
In recent years, the number of detected low-mass ($\la 0.20\;M_{\odot}$) helium white dwarfs (He~WDs) has
increased dramatically, mainly as a result of multiple survey campaigns such as 
WASP, ELM, HVS, {\it Kepler}, and SDSS \citep{psc+06,rbk+10,bgkk05,bkak10,bka+13,sob+12,kba+12,hmg+13,mbh+14}.

The existence of low-mass He~WDs in close binaries with a radio millisecond pulsar (MSP), however, 
has been known for a few decades \citep[e.g.][and references therein]{vbjj05}.
 Several attempts have been made to calibrate WD cooling models for such systems on the basis of the spin-down 
properties of the MSP \citep[e.g.][]{asvp96,hp98,dsbh98,asb01,pach07}.
The idea is that the characteristic spin-down age of the MSP ($\tau_{\rm PSR}\equiv P/(2\dot{P})$, where 
$P$ is the spin period and $\dot{P}$ is the period derivative)
should be equivalent to the cooling age of the WD ($\tau_{\rm cool}$), assuming that the radio MSP is activated 
at the same time as the WD is formed, following an epoch of mass transfer in a low-mass X-ray binary (LMXB). 
Unfortunately, this method is highly problematic since $\tau_{\rm PSR}$ generally is  a poor true age estimator.
It can easily be incorrect by a factor of 10 or more \citep{ctk94,llfn95,tau12,tlk12}.
Determining the true age of MSPs, however, is important for studying their spin evolution and
constraining the physics of their previous recycling phase \citep{ltk+14}.

The discovery of the intriguing PSR~J1012+5307 \citep{nll+95} sparked an intense discussion 
about WD cooling ages and MSP birthrates \citep{llfn95} given that $\tau_{\rm PSR} > 20\;\tau_{\rm cool}$. 
Soon thereafter, it was suggested \citep{asvp96,dsbh98,vbjj05} that He~WDs with a mass $\la 0.20\;M_{\odot}$ avoid hydrogen shell flashes, 
whereby their relatively thick ($\sim 10^{-2}\;M_{\odot}$) hydrogen envelope remains intact, causing residual hydrogen shell
burning to continue on a very long timescale.  
Despite significant theoretical progress \citep[e.g.][and references therein]{amc13}, our understanding of the thermal evolution of (proto) He~WDs remains uncertain. 
In particular, a number of recent observations of apparently bloated WDs calls for an explanation. 

In this Letter, we study the formation of a large number of low-mass He~WDs by modelling close-orbit LMXBs. We carefully investigate the properties of the resulting proto-WDs
and follow their evolution until and beyond settling on the WD cooling track. Finally, we compare our results with observations.

\begin{figure*}[t]
\begin{center}
\mbox{\includegraphics[width=0.50\textwidth, angle=-90]{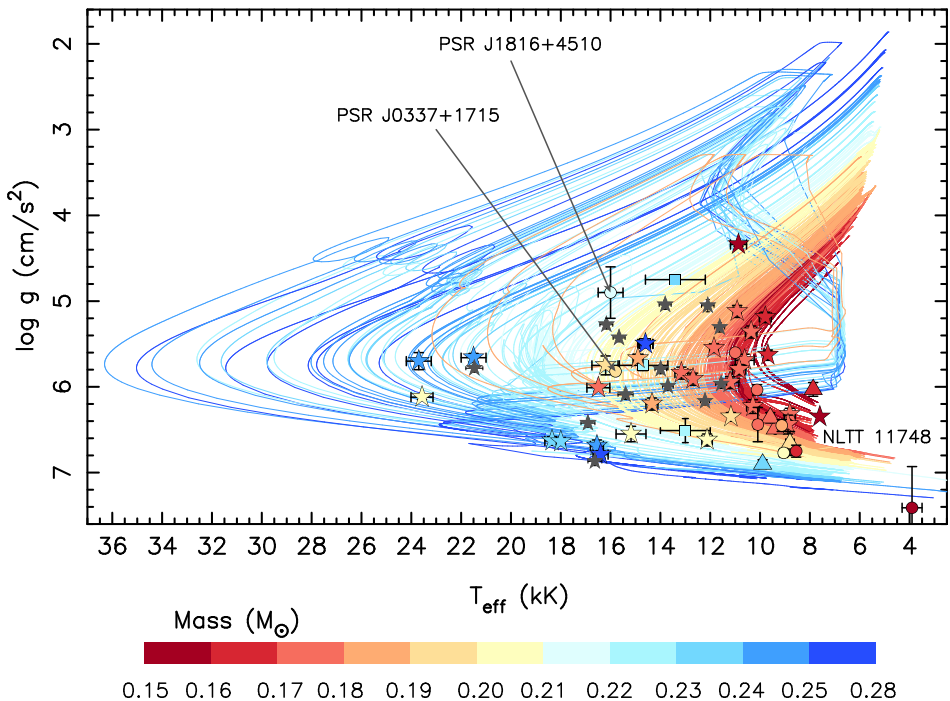}}
  \caption[]{
    Evolutionary tracks in the $(T_{\rm eff},\log\,g)$--diagram. The evolution from 
    Roche-lobe detachment until settling on the WD cooling track and beyond is shown for a selection of our models.  
    The colour scale represents the final WD mass.
    A few cases of vigorous hydrogen shell flashes explain the large (counterclockwise) loops in the diagram.  
    Observed WDs are shown with symbols (stars: sdB+WD, double WDs [grey stars: WDs with poor mass constraints]; triangles: pulsating WDs; squares: WD+MS; circles: WD+MSP 
    -- see Appendix~A for references and data). 
    }
\label{fig:logg-T}
\end{center}
\end{figure*}

\section{Numerical methods and physical assumptions }\label{sec:BEC}
Numerical calculations with a detailed stellar evolution code were used to trace the mass-transfer phase 
following the same prescription as outlined in \citet{itl14}.
We investigated models with a metallicity of $Z=0.02$, a mixing-length parameter $\alpha = l/H_p = 2.0$,  
and a core convective overshooting parameter of $\delta_{\rm OV}=0.10$.
A wide range of LMXB systems were investigated with different initial values of donor star mass ($M_2$), neutron star mass, orbital period,
and the so-called $\gamma$-index of magnetic braking. 
The evolution of the low-mass (proto) He~WD was calculated
including chemical diffusion (mixing), hydrogen shell flashes (CNO burning), and residual shell hydrogen burning. 
Convective, semi-convective, and overshoot mixing processes were treated via diffusion. 
Thermohaline mixing was included as well, whereas gravitational settling and radiative levitation were neglected, as was stellar wind mass loss.

\section{Results}\label{sec:Results}
In Fig.~\ref{fig:logg-T} we have plotted a selection of our calculated evolutionary tracks,  
from the moment of Roche-lobe detachment until the end of our calculations, 
for (proto) He~WDs with masses of $0.15-0.28\;M_{\odot}$. 
In general, our models fit the observations quite well. The few cases with discrepancies
are sources with large uncertainties in the WD mass. 
Vigorous single or multiple cycle hydrogen shell flashes explain the large loops in the diagram, whereas   
mild thermal instabilities are seen e.g. for the $0.25\;M_{\odot}$ proto-WDs at $\log g\simeq 4.5$.
It has been known for many years that 
a thermal runaway flash may develop through unstable CNO burning when a proto-WD evolves towards the cooling track \citep{kw67,web75,it86}. 
During these flashes the luminosity becomes very high, whereby the rate of hydrogen burning is significantly increased 
\citep[e.g.][and references therein]{ndm04,gau13}.
Our models with strong flashes often experience an additional episode of mass loss via Roche-lobe overflow \citep[RLO, see also][]{it86,seg00,prp02,ndm04}. 

For progenitor stars with $M_2 \le 1.5\;M_{\odot}$
we find hydrogen shell flashes in WDs with masses of $0.21 \le M_{\rm WD}/M_{\odot} \le 0.28$.
Hence, the lowest mass for which flashes occur is $M_{\rm flash}=0.21\;M_{\odot}$. 
However, we find a lower value of $M_{\rm flash}=0.18\;M_{\odot}$ for $M_2 = 1.6\;M_{\odot}$. 
It has been argued \citep[e.g.][]{vbjj05} that the value of $M_{\rm flash}$ is important since it marks 
a dichotomy for the subsequent WD cooling such that
WDs with a mass $M_{\rm WD} < M_{\rm flash}$ remain hot on a Gyr timescale as a result of continued residual 
hydrogen shell burning, whereas WDs with $M_{\rm WD} > M_{\rm flash}$ cool relatively fast as a result of the shell flashes that erode
the hydrogen envelope. We find that this transition is smooth, however, and that the thermal evolution timescale mainly depends 
 on the proto-He~WD luminosity and not on the occurrence or absence of flashes. 

\begin{figure}
\begin{center}
\mbox{\includegraphics[width=0.35\textwidth, angle=-90]{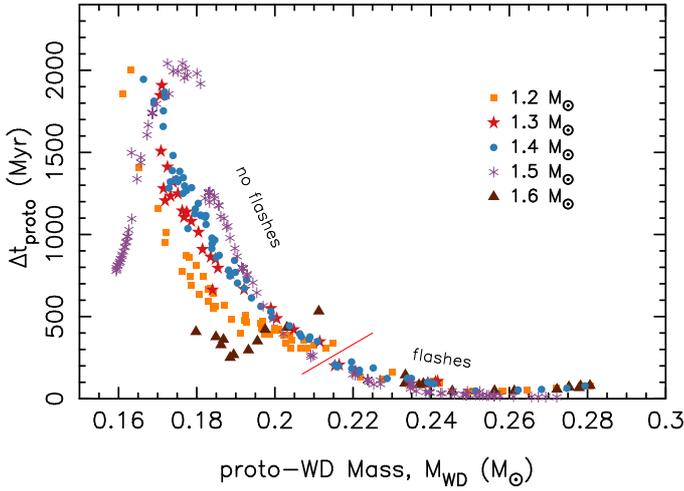}}
  \caption[]{
    Contraction timescale, $\Delta t_{\rm proto}$ of evolution from Roche-lobe detachment until settling on the WD cooling track,
    plotted as a function of WD mass, $M_{\rm WD}$.
    The initial ZAMS masses of the WD progenitors (the LMXB donor stars) are indicated with various symbols and colours.
    The red line marks $M_{\rm flash}\simeq 0.21\;M_{\odot}$ for progenitor stars $\la 1.5\;M_{\odot}$.
    }
\label{fig:delta-t}
\end{center}
\end{figure}

In Fig.~\ref{fig:delta-t} we have plotted the time, $\Delta t_{\rm proto}$ it takes from Roche-lobe detachment 
until the star reaches its highest value of $T_{\rm eff}$. 
(For WDs that undergo hydrogen shell flashes we used the time until the occurrence of highest $T_{\rm eff}$
on their last loop in the HR--diagram.) 
The plot shows a very strong dependence on $M_{\rm WD}$. For very low-mass He~WDs
(i.e. those with $M_{\rm WD}<M_{\rm flash}$, which therefore avoid hydrogen shell flashes), 
$\Delta t_{\rm proto}$ may last up to 2~Gyr. 
This result has important consequences for their thermal evolution and contraction (see below).
There is a well-known correlation between the degenerate core mass of an evolved low-mass star and its luminosity, $L$ \citep{rw71}.
After terminating the RLO, the star moves to the far left in the HR--diagram -- (initially) roughly at constant $L$ --
while burning the residual hydrogen in the envelope at a rate proportional to $L$. 
We find that the total amount of hydrogen left in the envelope is always $\sim\!0.01\pm0.005\;M_{\odot}$,
in agreement with \citet{seg00}, and is correlated in a variable manner with $M_{\rm WD}$ (especially for $M_2\ge1.5\;M_{\odot}$,
explaining the peak in Fig.~\ref{fig:delta-t}). 
Therefore, the increase in $\Delta t_{\rm proto}$ seen in Fig.~\ref{fig:delta-t} for decreasing values of $M_{\rm WD}$
can simply be understood from their much lower luminosities following the Roche-lobe detachment \citep[see also Figs.~5 and 10 in][]{itl14}. 
Based on our calculated proto-He~WD models, we find (see Appendix~B)
\begin{equation}
  \Delta t_{\rm proto} \simeq 400 \;\;{\rm Myr}\;\;\left(\frac{0.20\;M_{\odot}}{M_{\rm WD}}\right)^7 .
\label{eq:t_proto}
\end{equation}
 
The conclusion that $\Delta t_{\rm proto}$ can reach $\sim\!{\rm Gyr}$ was found previously for a few single models \citep[e.g.][]{dsbh98,seg00,asb01}.
Here we show, for the first time, its systematic dependence on $M_{\rm WD}$.  

\begin{figure}
\begin{center}
\mbox{\includegraphics[width=0.48\textwidth, angle=0]{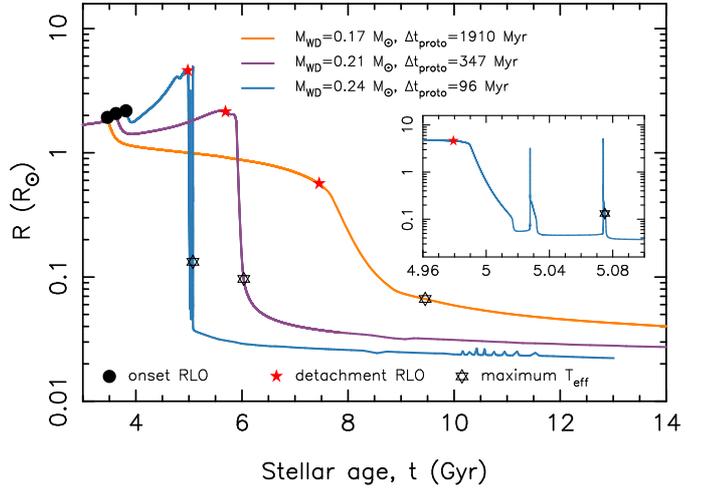}}
  \caption[]{
    Radius as a function of stellar age for the progenitor stars of three He~WDs of mass 0.17, 0.21 and $0.24\;M_{\odot}$.
    The most massive proto-WD evolves with hydrogen shell flashes -- see inlet.
    The epoch between the solid red star (RLO termination) and the open black star (max. $T_{\rm eff}$) marks
    the contraction (transition) timescale, $\Delta t_{\rm proto}=96-1910\;{\rm Myr}$.
    }
\label{fig:radius}
\end{center}
\end{figure}

Fig.~\ref{fig:radius} shows the contraction phase for three proto-He WDs. The value of $\Delta t_{\rm proto}$ increases significantly 
when $M_{\rm WD}$ decreases from 0.24 to $0.17\;M_{\odot}$. Hence, low-mass ($\la 0.20\;M_{\odot}$) proto-He WDs can remain
bloated on a very long timescale. It is important to notice that no pronounced discontinuity in $\Delta t_{\rm proto}$ is seen at $M_{\rm flash}\simeq 0.21\;M_{\odot}$
(cf. Figs.~\ref{fig:delta-t}, \ref{fig:radius}, and \ref{fig:t-proto-theo}). Although the peak luminosity (and thus the rate of eroding hydrogen) is high during a flash,
the star only spends a relatively short time ($\sim 10^3-10^6\;{\rm yr}$) at high $L$ when making a loop in the HR--diagram.

\section{Comparison  with observational data of He~WDs}\label{sec:obs}
In Table~\ref{table:obs} (Appendix~A) we list observed low-mass He~WDs included among the plotted data in Fig.~\ref{fig:logg-T}.
 We now discuss recent interesting sources in view of our theoretical modelling. 

\subsection{MSPs with low-mass (proto) He~WDs in tight orbits}\label{subsec:MSPs}
The companion star to PSR~J1816+4510 ($P_{\rm orb}=8.7\;{\rm hr}$) was recently observed by \citet{ksr+12,kbv+13}. They assembled optical spectroscopy and found 
an effective temperature of $T_{\rm eff}=16\,000\pm500\;{\rm K}$, a surface gravity of $\log g=4.9\pm0.3$, and a 
companion mass of $M_{\rm WD}\,\sin ^3i=0.193\pm 0.012 \;M_{\odot}$, where $i$ is the orbital inclination angle
of the binary. 
They concluded that while the spectrum is rather similar to that of a low-mass He~WD, it has a much lower surface gravity (i.e. larger radius) than a WD on the cooling track. 
They discussed that PSR~J1816+4510 perhaps represents a redback system \citep[cf.][for a formation scenario]{ccth13} where
pulsar irradiation of the hydrogen-rich, bloated companion causes evaporation of material, which can explain the observed eclipses of the radio pulses for $\sim\!$10\% of the orbit. 
However, the very hot surface temperature
of this companion (16$\,$000~K) cannot be explained from a redback scenario. 
Redbacks typically have illuminated dayside temperatures of only $T_{\rm eff}\simeq 6\,000\;{\rm K}$ \citep{bvr+13}.
Here we suggest that this companion star is simply a 
low-mass proto-He~WD. As we have demonstrated, such a star takes several $100\;{\rm Myr}$ to reach the cooling track, 
and our models match the observed values of $T_{\rm eff}$ and $\log g$. \citep[Note, for $P_{\rm orb}=8.7\;{\rm hr}$ 
one usually expects $M_{\rm WD}\la0.18\;M_{\odot}$, cf.][]{itl14}.

Another case is the triple system PSR~J0337+1715 recently discovered by \citet{rsa+14}, which raises fundamental questions about its formation \citep{tv14}. 
One open question is the order of formation of the two WDs orbiting the MSP (with $P_{\rm orb}=1.6$ and 327~days).
Spectroscopy of the inner companion by \citet{kvk+14} verified that this is a $0.197\;M_{\odot}$ He~WD, as known from pulsar timing. 
They measured a low surface gravity of $\log g = 5.82\pm0.05$ and noted that its very 
high surface temperature, $T_{\rm eff}=15,800\pm100\;{\rm K}$, could indicate that it had just experienced a flash. This would suggest a surprisingly short lifetime for this object. 
However, our modelling of $\sim\!0.20\;M_{\odot}$ He~WDs shows that these stars avoid flashes. 
Instead we find that for such a star it takes $400-600\;{\rm Myr}$ (Fig.~\ref{fig:delta-t}) to reach the WD cooling track.
Therefore, we conclude that it is reasonable to detect such a WD at an early, bloated stage of its evolution. 

\subsection{NLTT~11748 and other low-mass (proto) He~WD binaries}\label{subsec:11748}
A large number of low-mass proto-He~WDs (also classified as sdB~stars) are found in binaries with another WD. These systems probably formed via 
stable RLO in cataclysmic variable systems resembling our calculations, but with a $\sim\!0.7\;M_{\odot}$ CO~WD accretor instead of a NS. 
NLTT~11748 was discovered by \citet{sks+10}, with follow-up observations made by \citet{kmw+14}. 
This eclipsing detached binary consists of a $0.71-0.74\;M_{\odot}$ CO~WD with
a very low-mass He~WD and $P_{\rm orb}\simeq 5.6\;{\rm hr}$. Our evolutionary tracks for a $0.16\;M_{\odot}$ He~WD are indeed consistent with 
their observed values of $\log g = 6.35$ and $T_{\rm eff}=7\,600~{\rm K}$
(and their estimated mass of $0.136-0.162\;M_{\odot}$).
 \citet{bka+13} recently detected four binaries with low-mass WDs having $\log g\simeq 5$,
in accordance with our modelling of proto-He~WDs presented here (cf. Figs.~\ref{fig:logg-T} and ~\ref{fig:logg-T-time}).  

\subsection{Bloated, hot, low-mass He~WDs detected by {\it Kepler}}
Four (proto) He~WDs have been found with A-star companions
in the combined transit and eclipse data from the {\it Kepler} mission \citep{vrb+10,crf11,brvc12}. 
Three of these He~WDs (KOI--74, KIC~10657664, KOI--1224) have $M_{\rm WD}\la 0.26\;M_{\odot}$ and are also plotted in Fig.~\ref{fig:logg-T}. 
The mass estimates of these WDs are not very precise. However, within 1--2\,$\sigma$,  
the characteristics of these objects also seem to match our evolutionary tracks reasonably well.\\
The question now is  why we see all these bloated proto-WDs given that WDs spend significantly more time on the subsequent cooling tracks.
This is simply a selection effect. The WDs are only seen to eclipse A-stars in the {\it Kepler} data as long as they are bloated proto-WDs 
(and thus also more luminous than ordinary WDs, which have already settled on the cooling track).


\section{Discussion and conclusions}\label{sec:discussions}
We have demonstrated that low-mass ($\la 0.20\;M_{\odot}$), detached proto-He~WDs may spend up to $\sim\!2\;{\rm Gyr}$ 
in the contraction (transition) phase from the Roche-lobe detachment until they reach the WD cooling track. 
This is important for an age determination of He~WDs in general, and for recycled MSPs in particular. We expect a fair number of He~WDs to be observed in this (bloated) phase,
in agreement with recent observations.  

 The duration of this contraction phase ($\Delta t_{\rm proto}$) decreases strongly with increasing
mass of the proto-He~WD, $M_{\rm WD}$. This can be understood from the well-known correlation between
degenerate core mass and luminosity of an evolved low-mass star. Therefore, after Roche-lobe detachment, the rate at which
the residual ($0.01\pm0.005\;M_{\odot}$) hydrogen in the envelope is consumed is directly proportional to the luminosity 
and thus $M_{\rm WD}$. The value of $\Delta t_{\rm proto}$ is not particularly sensitive to the occurrence or absence of flashes.

Whether or not hydrogen shell flashes occur depends on the WD mass, its chemical composition, and the treatment of diffusion
(mixing) of the chemical elements \citep[e.g.][]{dsbh98,seg00,asb01,ndm04,amc13}.
In general, we find flashes in our models with $0.21\le M_{\rm WD}/M_{\odot} \le 0.28$ for $M_2\le 1.5\;M_{\odot}$.
This result agrees excellently well agreement with the interval found by \citet{ndm04} for donors with solar metallicity, and also with the earlier work of \citet{dsbh98}. For $M_2=1.6\;M_{\odot}$ we find that WDs down to $\sim\!0.18\;M_{\odot}$ are experiencing flashes.

Detailed studies by \citet{asb01,amc13} found hydrogen shell flashes 
for a much broader range of final WD mass ($0.18 < M_{\rm WD}/M_{\odot} < 0.41$). 
However, as pointed out by \citet{ndm04}, diffusion is an extremely fragile process, 
and turbulence can mitigate its effects. And more importantly, \citet{ndm04}  find that both $M_2$ and the mode of
angular momentum losses may also affect the range for which hydrogen shell flashes occur.
Indeed, we found a lower value of $M_{\rm flash}=0.18\;M_{\odot}$ for our models with $M_2 = 1.6\;M_{\odot}$.
It has previously been shown that $M_{\rm flash}$ strongly increases with lower metallicity \citep[e.g.][]{seg00,ndm04}.
The work of \citet{amc13} was calculated for a constant $M_2=1.0\;M_{\odot}$ ($Z=0.01$).
We have excluded such models with $M_2< 1.1\;M_{\odot}$ ($Z=0.02$) since these progenitor stars do not detach from their 
LMXB and evolve onto WD cooling track within a Hubble time.

Chemical diffusion via gravitational settling and radiative levitation was not included in this work. 
These effects seem to slightly increase $\Delta t_{\rm proto}$ compared with models without diffusion (L.~Nelson et al., in prep.). 
A systematic investigation of these and other effects on $\Delta t_{\rm proto}$ and $M_{\rm flash}$ will be addressed in a future work.

\begin{acknowledgements}
AGI acknowledges discussions with L.~Nelson, P.~Marchant, R.~Stancliffe and L.~Grassitelli. 
JA acknowledges financial support by the ERC Starting Grant no. (279702, BEACON, led by P.~Freire).
\end{acknowledgements}

\bibliographystyle{aa}
\bibliography{tauris_refs}

\Online

\begin{appendix} 
\section{Observational data and time evolution in the ($T_{\rm eff},\log g$)--diagram}\label{appendix:A}
\begin{figure}[]
\begin{center}
\mbox{\includegraphics[width=0.35\textwidth, angle=-90]{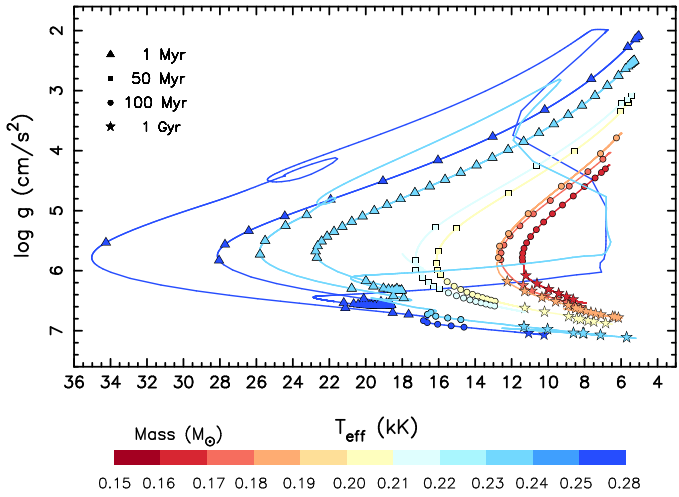}}
 \caption[]{
   Selected tracks (Fig.~\ref{fig:logg-T}) with a point marked for a time interval of 1~Myr (triangle),
   50~Myr (square), 100~Myr (circle), and 1~Gyr (star).
   }
\label{fig:logg-T-time}
\end{center}
\end{figure}

In Fig.~\ref{fig:logg-T-time} we have plotted points for fixed time intervals of evolution along a number of selected tracks from 
Fig.~\ref{fig:logg-T}. The density of points along these curves combined with the (proto) WD luminosities at these epochs can be used to
evaluate the probability of detecting them. For a direct comparison with data population synthesis needs to be included to probe the distribution of WD masses.
The observational data plotted in Fig.~\ref{fig:logg-T} were taken partly 
from the sources given in Table~\ref{table:obs} (primarily He~WDs with MSP companions, main-sequence A-star companions, or He~WDs
that have been detected to show pulsations). Additional data for the plotted symbols can be found in \citet{sob+12,hmg+13,bka+13}. 

\begin{table*}[]
\caption{Observational data of a number of low-mass He~WDs (and proto-He~WDs), preferentially in tight binary systems. }
\begin{tabular}{llllrl}
\hline  & & & & & \\ 
 He~WD                 & $\log g\;({\rm cm\,s}^{-2})$  & $T_{\rm eff}$ (K)  & $M_{\rm WD}\;(M_{\odot})$ & $P_{\rm orb}$ (hr) & Optical data\\
\noalign{\smallskip}
\hline 
\noalign{\smallskip}
 PSR~J0337+1715        & $5.82\pm0.05$                 & $15\,800\pm100$    & $0.197\pm 0.0002$  &  39.12     &  \citet{kvk+14}\\  
 PSR~J0348+0432        & $6.035\pm0.06$                & $10\,120\pm90$     & $0.172\pm0.003$   & 2.46       &  \citet{afw+13}\\  
 PSR~J0751+1807        & $7.41\pm0.48$                 & $\;\;3\,900\pm400$ & $0.138\pm 0.0006^a$ & 6.31     &  \citet{bvk06}\\  
 PSR~J1012+5307        & $6.75\pm0.07$                 & $\;\;8\,550\pm25$  & $0.16\pm0.02$      &  14.51     &  \citet{vbk96,cgk98}\\  
 PSR~J1738+0333        & $6.45\pm 0.07$                & $\;\;9\,130\pm150$ & $0.182\pm 0.016$   &   8.52     &  \citet{avk+12}\\  
 PSR~J1816+4510        & $4.9\pm 0.3$                  & $16\,000\pm500$    & $\sim\!0.21\pm 0.02^b$ & 8.66   &  \citet{ksr+12,kbv+13}\\  
 PSR~J1909$-$3744      & $6.77\pm 0.04$                & $\;\;9\,050\pm50$  & $0.2038\pm 0.0022$ &  36.72      &  \citet{ant13}\\
 PSR~J0024$-$7204U$^c$ & $\sim\!5.6$                   & $\sim\!11\,000$    & $\sim\!0.17$ &      10.29        &  \citet{egh+01}\\  
 PSR~J1911$-$5958A$^c$ & $6.44\pm 0.20$                & $10\,090\pm150$    & $0.175\pm 0.010$ &  20.64       &  \citet{bvkv06}\\  
\hline
\noalign{\smallskip}
 NLTT~11748            & $6.35\pm 0.03$                & $\;\;7\,600\pm120$ & $0.149\pm 0.013$ &  5.64        &  \citet{kmw+14}\\  
\hline
\noalign{\smallskip}
 KOI$-$74              & $6.51\pm 0.14$                & $13\,000\pm1000$   & $0.22\pm 0.03$ &  125.53          &  \citet{vrb+10}\\  
 KOI$-$1224            & $5.75\pm 0.06$                & $14\,700\pm1000$   & $0.22\pm 0.02$ &  64.75      &  \citet{brvc12}\\  
 KIC~10657664          & $5.50\pm 0.02$                & $14\,600\pm300$    & $0.26\pm 0.04^d$ & 78.55         &  \citet{crf11}\\  
\hline
\noalign{\smallskip}
 SDSS~J184037.78       & $6.49\pm 0.06$                & $\;\;9\,390\pm140$ & $\sim\!0.17$ &  4.59             &  \citet{hmg+13}\bf{$^{e,f}$}\\  
 SDSS~J111215.82       & $6.36\pm 0.06$                & $\;\;9\,590\pm140$ & $\sim\!0.17$ &  4.14           &  \citet{hmg+13}\bf{$^{e,f}$}\\  
 SDSS~J151826.68       & $6.90\pm 0.05$                & $\;\;9\,900\pm140$ & $\sim\!0.23$ & 14.62             &  \citet{hmg+13}\bf{$^{e,f}$}\\  
 J1614                 & $6.66\pm 0.14$                & $\;\;8\,800\pm170$ & $\sim\!0.19$ &  --          &  \citet{hmg+13}\bf{$^f$}\\  
 J2228                 & $6.03\pm 0.08$                & $\;\;7\,870\pm120$ & $\sim\!0.16$ &   --        &  \citet{hmg+13}\bf{$^f$}\\  
\hline
\noalign{\smallskip}
\end{tabular} 
\begin{flushleft}
 $^a$ D.~Nice, private comm. (2014).\\
 $^b$ Based on \citet{kbv+13}. See also \citet{itl14} for further comments on the component masses of this source.\\
 $^c$ The WD is most likely to have formed in this globular cluster binary given that the eccentricity is $e<10^{-5}$, as expected from recycling.\\
 $^d$  \citet{crf11} found two possible solutions for $M_{\rm WD}$ ($0.26\;M_{\odot}$ and $0.37\;M_{\odot}$). This WD has
      $P_{\rm orb}=3.3\;{\rm days}$ and thus we adopt the lower value of $M_{\rm WD}$ since this is agrees much better with the
      known $(M_{\rm WD},P_{\rm orb}$)-correlation \citep[see][for discussions]{ts99}. \\     
 $^e$ See additional references therein.\\
 $^f$ Pulsating He~WDs, see \citet{ca14} for recent theoretical modelling.

\end{flushleft}
\label{table:obs}
\end{table*}

\section{The (proto) WD contraction phase}\label{appendix:B}
\begin{figure}[]
\begin{center}
\mbox{\includegraphics[width=0.33\textwidth, angle=-90]{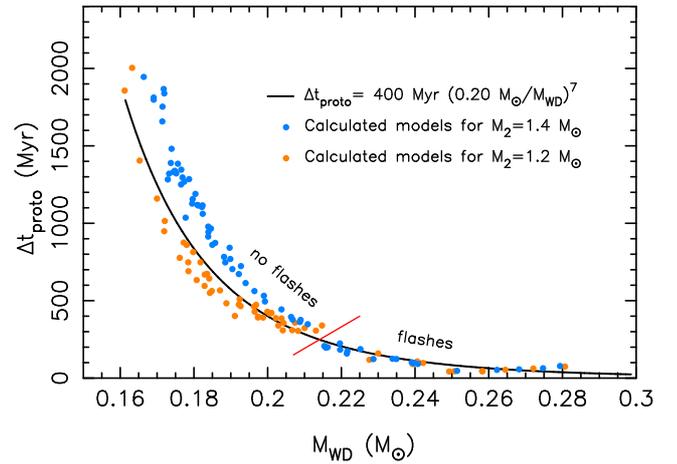}}
 \caption[]{
   Calculated models of proto-He~WDs from Fig.~\ref{fig:delta-t} for $M_2=1.2\;M_{\odot}$ (orange) and  $M_2=1.4\;M_{\odot}$ (blue).
   The black line is a fit to the data. It can also be derived analytically using a modified core mass--luminosity relation for
   low-mass evolved stars, combined with an assumed fixed amount of residual hydrogen ($0.01\;M_{\odot}$) to be burned. 
   The red line separates models with and without flashes.
   }
\label{fig:t-proto-theo}
\end{center}
\end{figure}

Fig.~\ref{fig:t-proto-theo} shows the time $\Delta t_{\rm proto}$
it takes from Roche-lobe detachment until the proto-He~WD reaches its highest value of $T_{\rm eff}$
and settles on the cooling track. Shown in this plot are all our calculated models for 
progenitor stars of 1.2 and $1.4\;M_{\odot}$ (i.e. a subset of the models plotted in Fig.~\ref{fig:delta-t}).
The black line (Eqn.~\ref{eq:t_proto}) is an analytical result obtained from a somewhat steep core mass--luminosity
function ($L\propto M_{\rm WD}^{\,7}$) combined with the assumption (for simplicity) that in all cases
$0.01\;M_{\odot}$ of hydrogen is burned before reaching the highest $T_{\rm eff}$.
The figure shows that this line also serves as a good approximate fit to our calculated models. 
For a given He~WD mass, the fit to $\Delta t_{\rm proto}$ calculated from our models is accurate to within 50\%.

\section{Nuclear burning during flashes}\label{appendix:C}
\begin{figure}[]
\begin{center}
\mbox{\includegraphics[width=0.32\textwidth, angle=-90]{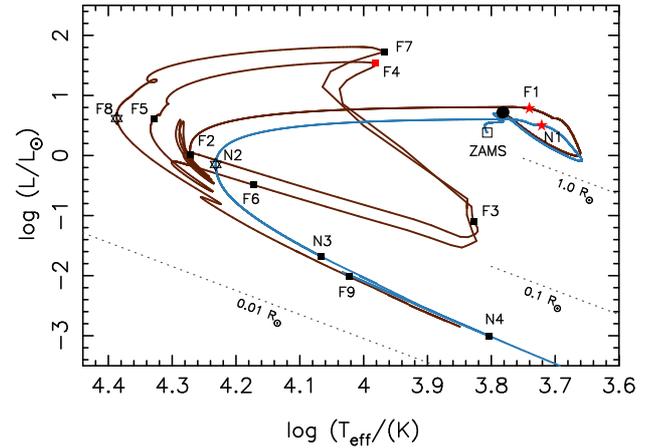}}
 \caption[]{Evolutionary tracks in the HR--diagram for a $0.221\;M_{\odot}$ proto-He~WD {\it with} flashes (brown) and 
            for a $0.212\;M_{\odot}$ proto-He~WD {\it without} flashes (blue).
            See Table~\ref{table:HR-flash} for data. 
   }
\label{fig:HR-flash_appendix}
\end{center}
\end{figure}

To compare the burning of residual envelope hydrogen for a case with and without large thermal instabilities (hydrogen shell flashes), we have plotted tracks in the HR--diagram
shown in Fig.~\ref{fig:HR-flash_appendix}. The age of the stars and the total amount of hydrogen remaining in their envelopes  
are given in Table~\ref{table:HR-flash} for the points marked in the figure.
These models were chosen very close to (but on each side of) $M_{\rm flash}\simeq 0.21\;M_{\odot}$, in both
cases for a $1.3\;M_{\odot}$ progenitor star.
As discussed in the main text, although the peak luminosity is high during a flash (and thereby the rate at
which hydrogen is burned), the star only spends
a relatively short time ($\sim\!10^6\;{\rm yr}$) in this epoch. 
(For more massive He~WDs it is even less time -- for example, it only lasts $\sim\!10^3{\rm yr}$ for a $0.27\;M_{\odot}$ He~WD.)
Therefore, the amount of additional hydrogen burned as a result of flashes is relatively small. In the example shown in
Fig.~\ref{fig:HR-flash_appendix} it amounts to about 12\% of the total amount of hydrogen at the point of Roche-lobe detachment.
Hence, the flashes may appear to reduce $\Delta t_{\rm proto}$ by $\sim\!100\;{\rm Myr}$. However, one must
bear in mind that the  proto-WDs that experience flashes are also the WDs with the least amount of hydrogen
in their envelopes after RLO. 

For a star that experiences flashes, the residual hydrogen present in the envelope following the LMXB-phase
is processed roughly as follows: 70\% during the epoch from Roche-lobe detachment
until reaching highest $T_{\rm eff}$, 10\% during the flashes, and 20\% after finally settling on the WD cooling track. 

\begin{table*}[]
\caption{Ages and remaining hydrogen of the two proto-He~WDs shown in Fig.~\ref{fig:HR-flash_appendix}. 
         The points N1...N4 are for the non-flashing $0.212\;M_{\odot}$ WD model and the points F1...F9 are for the $0.221\;M_{\odot}$ flashing WD model.}
\begin{tabular}{rrrr}
\hline  & & & \\ 
                     Point & Relative age$^a$       & Total age$^b$       & Hydrogen$^c$ \\
                           &                        &                     & $(10^{-3}\;M_{\odot})$ \\
\noalign{\smallskip}
\hline 
\noalign{\smallskip}
                      N1   &        0               &          0          & 13.68 \\
                      N2   &      341 Myr           &        341 Myr      &  2.94 \\
                      N3   &   1\,900 Myr           &     2\,240 Myr      &  0.79 \\
                      N4   &   8\,231 Myr           &    10\,470 Myr      &  0.67 \\
\noalign{\smallskip}
\hline 
\noalign{\smallskip}
                      F1   &        0               &          0          &  7.78 \\
                      F2   &      107 Myr           &        107 Myr      &  2.71 \\
                      F3   &       31 Myr           &        138 Myr      &  2.45 \\
                      F4   &     1536 yr            &        138 Myr      &  2.45 \\
                      F5   &      5.1 Myr           &        143 Myr      &  2.22 \\
                      F6   &       20 Myr           &        163 Myr      &  2.06 \\
                      F7   &     1536 yr            &        163 Myr      &  2.05 \\
                      F8   &      3.6 Myr           &        166 Myr      &  1.75 \\
                      F9   &   2\,089 Myr           &     2\,255 Myr      &  0.85 \\
\hline
\noalign{\smallskip}
\end{tabular} 
\begin{flushleft}
 $^a$ Age relative to the previous point along the track.\\
 $^b$ Cumulated age relative to the first point on the track (since the time of Roche-lobe detachment).\\
 $^c$ Total amount of hydrogen remaining in the envelope of the (proto)-He~WD.\\
\end{flushleft}
\label{table:HR-flash}
\end{table*}

\end{appendix}

\end{document}